\documentclass[final,5p,twocolumn,times,sort&
compress]{elsarticle}

\usepackage{mathrsfs}
\usepackage[latin1]{inputenc}
\usepackage{array}
\usepackage{dsfont}
\usepackage{amsmath}\usepackage{amssymb,stmaryrd}
\usepackage{booktabs}
\usepackage{lipsum}
\usepackage{cancel}
\usepackage{hyperref}

\def\preprintdate{March 2019}

\renewcommand{\eqref}[1]{\mbox{Eq.~(\ref{#1})}}
\newcommand{\tabref}[1]{\mbox{Tab.~\ref{#1}}}

\newcommand{\secref}[1]{\mbox{Sec.~\ref{#1}}}

\begin{document}

\begin{frontmatter}

\title{Classical Lagrangians for the nonminimal Standard-Model Extension \\ at higher orders in Lorentz violation}

\author{Marco Schreck}

\address{Departamento de F\'{\i}sica, Universidade Federal do Maranh\~{a}o (UFMA) \\
Campus Universit\'{a}rio do Bacanga, S\~{a}o Lu\'{\i}s -- MA, 65085-580, Brazil}

\address{}
\address{\rm
\preprintdate
}

\begin{abstract}

The current paper is dedicated to determining perturbative expansions for Lagrangians describing classical, relativistic, pointlike particles subject to Lorentz violation parameterized by the nonminimal Standard-Model Extension (SME). An iterative technique recently developed and applied to a Lorentz-violating scalar field theory is now adopted to treat the spin-degenerate SME fermion sector. Lagrangians are obtained at third order in Lorentz violation for the operators $\hat{a}_{\mu}$, $\hat{c}_{\mu}$, $\hat{e}$, $\hat{f}$, and $\hat{m}$ for arbitrary mass dimension. The results demonstrate the impact of nonzero spin on classical particle propagation. They will be useful for phenomenological studies of modified gravity and could provide useful insights into explicit Lorentz violation in curved spacetimes.

\end{abstract}

\end{frontmatter}

\section{Introduction}

The quest for a violation of Lorentz invariance in nature continues. Minuscule violations of this fundamental symmetry could be detected as a remainder of Planck-scale quantum gravity effects such as strings \cite{Kostelecky:1988zi}, spin foam in the context of loop quantum gravity \cite{Gambini:1998it}, noncommutative spacetimes \cite{AmelinoCamelia:1999pm}, spacetime foam \cite{Klinkhamer:2003ec}, nontrivial spacetime topologies \cite{Klinkhamer:1998fa}, and effects from possible UV-completions of general relativity such as Ho\v{r}ava-Lifshitz gravity \cite{Horava:2009uw}. As there is still no theory of quantum gravity at hand, broad and competitive experimental searches for Lorentz violation are recommended. Such searches are best carried out in a general framework that incorporates all possible deviations from Lorentz invariance. The Standard-Model Extension (SME) provides such a framework as an effective field theory parameterizing Lorentz violation for energies much smaller than the Planck energy. A generic Lorentz-violating contribution consists of a tensor-valued background field contracted with a suitable field operator to form a scalar under coordinate changes (observer Lorentz transformations). The minimal SME \cite{Colladay:1996iz} comprises all contributions contracted with field operators of mass dimensions 3 and 4, whereas the nonminimal SME \cite{Kostelecky:2009zp,Kostelecky:2013rta} includes field operators of arbitrary mass dimension $d$. A subset of the Lorentz-violating operators also violates discrete \textit{CPT} symmetry, i.e., \textit{CPT} violation at the level of effective field-theory is automatically contained in the SME \cite{Greenberg:2002uu}.

In the community, a certain interest has developed on descriptions of classical, relativistic, pointlike particles subject to the Lorentz-violating background fields of the SME. The connection was first established in \cite{Kostelecky:2010hs} triggering a large series of related works \cite{LagFinslerPapers} in the aftermath. As the SME is based on field theory, such a link is not obvious at first, but it is highly useful for several reasons. First, there are phenomenological motivations. Experiments testing general relativity are usually carried out with macroscopic test masses, whereby quantum effects are negligible in this regime. Hence, it is highly desirable to be able to describe the motion of such classical bodies under the influence of Lorentz violation parameterized by the SME. Second, a classical description of Lorentz violation is also interesting from a theoretical and even mathematical point of view. It was shown that a classical particle subject to Lorentz violation moves along a trajectory in a Riemann-Finsler space \cite{Kostelecky:2011qz}. Therefore, Lorentz violation has a direct connection to Finsler geometry that is based on path length functionals beyond the quadratic restriction of Riemannian geometry \cite{Riemann:2004,Finsler:1918}. Finsler geometry is a promising candidate for the development of a better understanding of explicit Lorentz violation in curved spacetimes.

This article can be considered as a follow-up of the recent work \cite{Edwards:2018lsn}. In the latter paper, a perturbative iterative procedure was developed to derive a classical Lagrangian successively from the modified dispersion equation connected to the field theory description. This procedure was successfully applied to a Lorentz-violating modification of a scalar field theory. The next reasonable step is to take spin into account, i.e., to apply the method to a field theory of modified Dirac fermions. The goal of the current paper is to carry out this task for spin-degenerate Lorentz violation. It is organized as follows. Section~\ref{sec:dirac-theory} provides a summary of the most essential aspects of the modified SME fermion sector that are of relevance to this work. The first part of \secref{sec:classical-kinematics} is dedicated to a brief summary of the problem of classical Lagrangians. In the second part, the method introduced in~\cite{Edwards:2018lsn} is reviewed and applied to spin-degenerate Lorentz violation for fermions. Finally, the results are stated in \secref{sec:third-order-lagrangians} with its properties to be discussed subsequently. Section~\ref{sec:conclusions} concludes on the findings and provides an outlook to further studies that could be worthwhile to be pursued in the future. Calculational details are relegated to the appendix. Natural units with $\hbar=c=1$ will be used unless otherwise stated. The notation employed in \cite{Edwards:2018lsn} will be largely taken over.

\section{Modified Dirac field theory}
\label{sec:dirac-theory}

The fermion sector of the nonminimal SME was introduced in \cite{Kostelecky:2013rta} where it was also studied extensively. It is based on a modified Dirac Lagrange density of the form
\begin{equation}
\mathcal{L}=\frac{1}{2}\overline{\psi}(\mathrm{i}\cancel{\partial}-m_{\psi}\mathds{1}_4+\hat{\mathcal{Q}})\psi+\text{H.c.}\,,
\end{equation}
with the Dirac field $\psi$, the Dirac conjugate field $\overline{\psi}\equiv\psi^{\dagger}\gamma^0$, the unit matrix $\mathds{1}_4$ in spinor space, and the fermion mass $m_{\psi}$. The fields are defined in Minkowski spacetime with metric tensor $\eta_{\mu\nu}$ of signature $(+,-,-,-)$. The standard Dirac matrizes satisfy the Clifford algebra $\{\gamma_{\mu},\gamma_{\nu}\}=2\eta_{\mu\nu}\mathds{1}_4$. The Lorentz-violating contributions are contained in the operator $\hat{\mathcal{Q}}$ whose Dirac structure is expressed in terms of the 16 Dirac matrices that appear in Dirac bilinears. Each set of matrices is contracted with a Lorentz-violating operator that is a contraction of controlling coefficients and four-derivatives on its part. There are two classes of such operators. The first comprises the spin-degenerate operators $\hat{a}_{\mu}$, $\hat{c}_{\mu}$, $\hat{e}$, $\hat{f}$, and $\hat{m}$ where the second involves the spin-nondegenerate ones $\hat{b}_{\mu}$, $\hat{d}_{\mu}$, $\hat{H}_{\mu\nu}$, and $\hat{g}_{\mu\nu}$. The spin-degenerate operators do not lift the two-fold degeneracy of positive-energy fermion dispersion relations. Hence, both spin-up and spin-down fermions are described by the same dispersion relation. In contrast, the spin-nondegenerate operators lead to distinct dispersion relations for fermions with spin-up and those with spin-down. The spin-degenerate operators only will be subject to the investigations to be carried out in the current paper. In momentum space, all four-derivatives are replaced by four-momenta via $p_{\mu}=\mathrm{i}\partial_{\mu}$.

In the context of field theory, physical fields can be redefined without changing physical observables. Such field redefinitions are known for the fermion sector connecting coefficients of different types to each other or removing them completely from the Lagrange density. For example, there is a field redefinition that eliminates the minimal $a$ coefficients, as long as gravity effects are not taken into account. However, once fermion fields of different flavors are present, $a^{(3)}_{\mu}$ cannot be removed simultaneously from all particle sectors.

Apart from the latter one, a further field redefinition is known that identifies the minimal $f$ coefficients (squared) with minimal $c$ coefficients \cite{Altschul:2006ts}. This means that the $f$ coefficients are not distinguishable from the $c$ coefficients and that they appear at least to quadratic order in observables. A similar map seems to exist in the nonminimal SME, but greater care has to be taken to match the number of Lorentz indices. Comparing the dispersion equations for the $\hat{c}_{\mu}$ and $\hat{f}$ operators with each other, the following analog correspondence can be derived within the nonminimal SME (see \ref{sec:map-c-f-coefficients} for details):
\begin{subequations}
\label{eq:correspondence-c-f}
\begin{align}
c^{(d)\mu\nu\diamond}&=\frac{(f^{(\overline{d})})^{\mu\diamond}(f^{(\overline{d})})^{\nu\diamond}}{\Theta}\left[\sqrt{1-\Theta}-1\right] \notag \\
&=-(f^{(\overline{d})})^{\mu\diamond}(f^{(\overline{d})})^{\nu\diamond}\left(\frac{1}{2}+\frac{\Theta}{8}+\frac{\Theta^2}{16}+\dots\right)\,, \displaybreak[0]\\[2ex]
\Theta&=(f^{(\overline{d})})_{\alpha}^{\phantom{\alpha}\diamond}(f^{(\overline{d})})^{\alpha\diamond}\,,\quad d=2(\overline{d}-2)\,,
\end{align}
\end{subequations}
with mass dimensions $d$, $\overline{d}$. Here, $\diamond$ is a convenient short-hand notation that describes the contraction of a coefficient with an appropriate number of four-momenta, e.g., $(a^{(d)})^{\diamond}\equiv (a^{(d)})^{\alpha\dots\alpha_{d-3}}p_{\alpha}\dots p_{\alpha_{d-3}}$. Note that the mass dimensions of the $c$ and $f$ coefficients differ from each other. For the minimal coefficients, we obtain $d=2(4-2)=4$ based on $\overline{d}=4$, which reproduces Eq.~(19) for \cite{Altschul:2006ts}. However, it is not possible to find a matching set of $f$ coefficients to each set of $c$ coefficients. This is the case when $d/2$ is an odd number, as the mass dimension of the $f$ coefficients is always even. For example, there are no $f$ coefficients corresponding to $c^{(6)}$.

We will consider all controlling coefficients as totally symmetric with respect to permutations of their Lorentz indices. For the parts that are contracted with four-derivatives this is justified, as these commute with each other. In principle, coefficients can have antisymmetric parts when they have Lorentz indices that are contracted with objects other than four-derivatives (for example, $\gamma^{\mu}$). As will become clear below, these antisymmetric pieces can contribute to the Lagrangian at quadratic and higher orders only. Working with totally symmetric coefficients greatly simplifies the calculations and covers a major part of the interesting cases.

\section{Classical kinematics}
\label{sec:classical-kinematics}

Round about a decade ago a procedure was developed to describe the kinematics of a classical, relativistic, pointlike particle that moves along the trajectory of the centroid of a quantum wave packet constructed from modified plane-wave solutions of the SME field theory \cite{Kostelecky:2010hs}. On the one hand, the wave packet is described by a modified dispersion equation $\mathcal{D}(p)=0$ that is a function of the four-momentum $p_{\mu}$. On the other hand, a classical particle is described by a Lagrangian $L(u)$ in terms of the four-velocity $u^{\mu}$. Note that the trajectory of the classical particle is parameterized arbitrarily, i.e., we do not explicitly use proper time such that $u^2=1$ does not necessarily hold. In the latter case, $u^{\mu}$ is not a four-vector and $L$ is not a Lorentz scalar, anymore. Arguments based on the vector character of $u^{\mu}$ must be understood to be valid for a proper-time parameterization of the trajectory.

The correspondence between the wave packet and the classical particle is established by a set of five ordinary, nonlinear equations that involve the four-momentum components, the four-velocity components, and the Lagrangian:
\begin{subequations}
\label{eq:mapping-equations}
\begin{align}
\label{eq:mapping-equation-1}
\mathcal{D}(p)&=0\,, \displaybreak[0]\\[1ex]
\label{eq:mapping-equation-2-3-4}
\frac{\partial p_0}{\partial p_i}&=-\frac{u^i}{u^0}\,,\quad i\in \{1,2,3\}\,, \displaybreak[0]\\[1ex]
\label{eq:mapping-equation-5}
L&=-p_{\mu}u^{\mu}\,.
\end{align}
\end{subequations}
Here, \eqref{eq:mapping-equation-1} is the dispersion equation that follows from the determinant of the modified Dirac operator or simply from Eq.~(14) of \cite{Kostelecky:2000mm} for the minimal SME and Eq.~(39) of \cite{Kostelecky:2013rta} for the nonminimal SME. Equations~(\ref{eq:mapping-equation-2-3-4}) state a correspondence between the group velocity of the quantum wave packet and the three-velocity of the pointlike particle. Finally, \eqref{eq:mapping-equation-5} is Euler's theorem valid for a Lagrangian that is positively homogenous of degree 1 in the velocity: $L(\lambda u)=\lambda L(u)$ for $\lambda>0$. The latter property guarantees that the action is invariant under reparameterizations of the classical trajectory that is a reasonable property to have from a physicist's perspective.  Note that this method works for massive fermions only, whereas for photons a different technique must be applied based on the eikonal approach in classical optics \cite{Schreck:2015dsa}.

For the spin-degenerate coefficients of the minimal SME, Eqs.~(\ref{eq:mapping-equations}) can be solved with standard methods resulting in Lagrangians describing the sectors spanned by the $a$, $e$, $f$ coefficients and the $c$ coefficients, respectively \cite{Kostelecky:2010hs}. In principle, it is even possible to construct $L_{\mathrm{face}}$ describing the full coefficient choice, although its form is probably not very transparent.

For most sets of nonminimal coefficients, four of the five equations involve high powers of four-momentum components making a direct solution challenging or even impossible.
Because of this, over a period of several years further methods had to be applied to solve this problem and to make such calculations practical in the context of the nonminimal SME:

\begin{enumerate}

\item Gr\"{o}bner bases to solve the system of polynomial equations~\cite{Schreck:2015seb}.
\item Ansatz-based technique for a Lagrangian and subsequent check of the system of equations \cite{Reis:2017ayl}.
\item Perturbative iterative solution of the nonlinear system of equations \cite{Edwards:2018lsn}.

\end{enumerate}

Applying the first two culminated in the full classical Lagrangian for the nonminimal SME at leading order in Lorentz violation \cite{Reis:2017ayl}. The third technique is the most recent one and was applied to a scalar field theory including nonminimal $a$- and $c$-type operators of arbitrary mass dimension. It works very well for dispersion equations of the generic form
\begin{subequations}
\label{eq:dispersion-equation-generic-minimal-spin-degenerate}
\begin{align}
\mathcal{D}(p)&=0\,, \displaybreak[0]\\[2ex]
\mathcal{D}(p)&=(p+\kappa)_{\mu}\Omega^{\mu\nu}(p+\kappa)_{\nu}-\mu^2\,.
\end{align}
\end{subequations}
Here, $\kappa^{\mu}=\kappa^{\mu}(p)$ is a shift of the four-momentum, $\mu=\mu(p)$ a Lorentz scalar, and $\Omega^{\mu\nu}=\Omega^{\mu\nu}(p)$ a symmetric $4\times 4$ matrix, which is a perturbation of the Minkowski metric tensor \cite{Kostelecky:2010hs}. In the Lorentz-invariant case we have that $\kappa^{\mu}=0^{\mu}$, $\Omega^{\mu\nu}=\eta^{\mu\nu}$, and $\mu=m_{\psi}$. These quantities are simply constants for the minimal SME, but they depend on the momentum in case of the nonminimal SME. The dispersion equation of each type of spin-degenerate operator is of the form
\begin{equation}
\mathcal{D}^2(p)=0\,,
\end{equation}
which is why the technique is adequate for this sector, as well. The steps of the procedure are to differentiate the dispersion equation for $p_i$, contract it with $p_i$, and use \eqref{eq:mapping-equation-5} and the dispersion equation again to obtain the four-velocity in terms of the four-momentum. Carrying these steps out for the spin-degenerate coefficients leads to the following results (cf.~\ref{sec:computation-four-velocities} for details):
\begin{subequations}
\label{eq:four-velocities-cases}
\begin{align}
u^{\mu}|_a&=-L\frac{p^{\mu}-(a^{(d)})^{\mu\diamond}-(d-3)(a^{(d)})^{\mu\kappa\diamond}\left(p_{\kappa}-(a^{(d)})_{\kappa}^{\phantom{\kappa}\diamond}\right)}{m_{\psi}^2-(d-4)(a^{(d)})^{\kappa\diamond}\left(p_{\kappa}-(a^{(d)})_{\kappa}^{\phantom{\kappa}\diamond}\right)}\,, \notag \displaybreak[0]\\[2ex]
\label{eq:four-velocity-a}
u^{\mu}|_c&=-L\frac{p^{\mu}+(d-2)(c^{(d)})^{\mu\diamond}+(d-3)(c^{(d)})^{\mu\kappa\diamond}(c^{(d)})_{\kappa}^{\phantom{\kappa}\diamond}}{m_{\psi}^2+(d-4)[(c^{(d)})^{\diamond}+(c^{(d)})^{\kappa\diamond}(c^{(d)})_{\kappa}^{\phantom{\kappa}\diamond}]}\,, \displaybreak[0]\\[2ex]
\label{eq:four-velocity-c}
u^{\mu}|_e&=-L\frac{p^{\mu}+(d-3)\left[m_{\psi}-(e^{(d)})^{\diamond}\right](e^{(d)})^{\mu\diamond}}{\left[m_{\psi}+(d-4)(e^{(d)})^{\diamond}\right]\left[m_{\psi}-(e^{(d)})^{\diamond}\right]}\,, \displaybreak[0]\\[2ex]
\label{eq:four-velocity-e}
u^{\mu}|_f&=-L\frac{p^{\mu}-(d-3)(f^{(d)})^{\mu\diamond}(f^{(d)})^{\diamond}}{m_{\psi}^2-(d-4)[(f^{(d)})^{\diamond}]^2}\,, \displaybreak[0]\\[2ex]
u^{\mu}|_m&=-L\frac{p^{\mu}-(d-3)\left[m_{\psi}+(m^{(d)})^{\diamond}\right](m^{(d)})^{\mu\diamond}}{\left[m_{\psi}-(d-4)(m^{(d)})^{\diamond}\right]\left[m_{\psi}+(m^{(d)})^{\diamond}\right]}\,.
\end{align}
\end{subequations}
The four-velocities of this form are to be contracted with $u_{\mu}$ where \eqref{eq:mapping-equation-5} is employed once more to obtain a quadratic polynomial in $L$. For example, for the operator $\hat{e}$:
\begin{subequations}
\begin{align}
0&=\zeta_eL_e^2+\psi_eL_e-u^2\,, \displaybreak[0]\\[2ex]
\zeta_e&=\frac{1}{\left[m_{\psi}+(d-4)(e^{(d)})^{\diamond}\right]\left[m_{\psi}-(e^{(d)})^{\diamond}\right]}\,, \displaybreak[0]\\[2ex]
\psi_e&=-\frac{(d-3)\left[m_{\psi}-(e^{(d)})^{\diamond}\right](e^{(d)})^{\mu\diamond}}{\left[m_{\psi}+(d-4)(e^{(d)})^{\diamond}\right]\left[m_{\psi}-(e^{(d)})^{\diamond}\right]}\,,
\end{align}
\end{subequations}
and analogously for the remaining ones. Solving these polynomials for $L$ provides Lagrangians that are only partially expressed in terms of the four-velocity, but they still involve the four-momentum, as well. It is a Lagrangian of this form that proves to be extremely useful for a perturbative, iterative treatment of the problem. For zero Lorentz violation, the standard Lagrangian $L^{(d)}_0=-m_{\psi}\overline{u}$ with $\overline{u}\equiv\sqrt{u^2}$ is readily obtained. The covariant momentum associated with this Lagrangian is given by $(p_0)_{\mu}\equiv-\partial L^{(d)}_0/\partial u^{\mu}$. Inserting this result into the Lagrangian and keeping terms to first order in Lorentz violation delivers $L^{(d)}_1$ and $(p_1)_{\mu}$ subsequently. Continuing this iteration, provides $L^{(d)}_{q+1}$ valid at $(q+1)$-th order in Lorentz violation from $L^{(d)}_q$ and $(p_q)_{\mu}\equiv-\partial L^{(d)}_q/\partial u^{\mu}$.
\begin{table*}[t]
\centering
\setlength\extrarowheight{5pt}
\begin{tabular}{ccccccc}
\toprule
Parameters & $\hat{a}^{(d)}_{\mu}$ & $\hat{e}^{(d)}$ \\
\midrule
$\xi^{(1)}_{1;a,e}$ & 1 & $-1$ \\
$\xi^{(2)}_{1;a,e}$ & $-\frac{1}{2}(d-3)^2$ & $-\frac{1}{2}(d-3)(d-5)$ \\
$\xi^{(2)}_{2;a,e}$ & $\frac{1}{2}(d-1)(d-3)$ & $\frac{1}{2}(d-3)^2$ \\
$\xi^{(3)}_{1;a,e}$ & $\frac{1}{2}(d-3)^4$ & $-\frac{1}{2}(d-3)(d-4)(d^2-9d+19)$ \\
$\xi^{(3)}_{2;a,e}$ & $-\frac{1}{2}(d-1)(d-3)^2(2d-7)$ & $\frac{1}{2}(d-3)^2(2d^2-17d+34)$ \\
$\xi^{(3)}_{3;a,e}$ & $\frac{1}{2}(d-2)(d-3)(d^2-4d+2)$ & $-\frac{1}{2}(d-3)^3(d-4)$ \\
\bottomrule
\end{tabular}
\caption{Parameters of the generic Lagrangian of \eqref{eq:lagrangian-expansion-generic} obtained at fixed $d$ for the spin-degenerate operators $\hat{a}_{\mu}$, $\hat{e}$ of the nonminimal SME fermion sector. Apart from the results stated, it holds that $\xi^{(m)}_{i;c}=(-1)^m\xi_{i;a}$ and $\xi^{(m)}_{i;m}=(-1)^m\xi_{i;e}$, i.e., these parameters are omitted explicitly.}
\label{tab:results-parameters-lagrangians-a-e}
\end{table*}
For an operator of fixed mass dimension $d$, the calculation is best carried out with computer algebra for a specific choice of a single, simple controlling coefficient. Following this procedure, it is relatively straightforward to express the result in terms of observer Lorentz scalars, vectors, and tensors that are constructed from the background field involved. Due to observer Lorentz invariance, the Lagrangian obtained for a single coefficient must be valid for the full operator. For arbitrary $d$ it turned out to be reasonable to restrict computations to $(1+1)$ spacetime dimensions to save time and resources. The computation is performed with a generic coefficient of mass dimension $d$ that is multiplied by a $p$-dependent power of the spatial momentum. The resulting Lagrangian is again expressed in terms of suitable observer Lorentz scalars, vectors, and tensors.

To be sure about the correctness of a particular Lagrangian, the associated conjugated momentum is tested numerically to satisfy the set of five nonlinear equations to the order under consideration. To do so, it is not even necessary to obtain the dispersion relation, which may be problematic both for a large number of controlling coefficients and coefficients that are contracted with additional $p_0$ components. Instead, the dispersion equation can be implicitly differentiated for $p_i$ with $\partial p_0/\partial p_i=-u^i/u^0$ to be employed afterwards. Inserting the correct Lagrangian at order $q$, should produce numerical errors of order $X^{q+1}$ for generic controlling coefficients $X$. Choosing minuscule numerical values for these coefficients (for example, of an order of magnitude of $10^{-10}$), the Lagrangian is supposedly correct when the numerical error is of the order $10^{-10(q+1)}$.

\section{Third-order classical Lagrangians}
\label{sec:third-order-lagrangians}

The iterative procedure described above was applied to obtain classical Lagrangians for the spin-degenerate nonminimal operators of the SME fermion sector with arbitrary mass dimension. The iteration was stopped after computing the contribution at third order in Lorentz violation for the operators $\hat{a}_{\mu}$, $\hat{c}_{\mu}$, $\hat{e}$, $\hat{m}$ and at sixth order for $\hat{f}$. Based on the choice of a single coefficient, such calculations are practical with computer algebra and do not require a large amount of time or resources. Hence, higher orders could, in principle, be computed. However, as all experiments to date have demonstrated that Lorentz violation in Minkowski spacetime must be minuscule, third-order expansions are expected to be sufficient for all practical purposes. The generic form of the Lagrangians for $\hat{a}_{\mu}$, $\hat{c}_{\mu}$, $\hat{e}$, and $\hat{m}$ is as follows:
\begin{subequations}
\label{eq:lagrangian-expansion-generic}
\begin{align}
L_{3,X}^{(d)}&=L_0^{(d)}\left[1+\xi_{1;X}^{(1)}\tilde{X}^{(d)}+\xi^{(2)}_{1;X}(\tilde{X}^{(d)})^2+\xi^{(2)}_{2;X}(\tilde{X}^{(d)})_{\alpha}(\tilde{X}^{(d)})^{\alpha}\right. \notag \displaybreak[0]\\
&\phantom{{}={}L_0^{(d)}\Big\{}\left.+\,\xi^{(3)}_{1;X}(\tilde{X}^{(d)})^3+\xi^{(3)}_{2;X}\tilde{X}^{(d)}(\tilde{X}^{(d)})_{\alpha}(\tilde{X}^{(d)})^{\alpha}\right. \notag \displaybreak[0]\\
&\phantom{{}={}L_0^{(d)}\Big\{}\left.+\,\xi^{(3)}_{3;X}(\tilde{X}^{(d)})_{\alpha}(\tilde{X}^{(d)})^{\alpha\beta}(\tilde{X}^{(d)})_{\beta}\right]\,,
\end{align}
with contractions of generic controlling coefficients and four-velocities
\begin{equation}
(\tilde{X}^{(d)})_{\alpha_1\dots\alpha_l}=m_{\psi}^{d-4}(X^{(d)})_{\alpha_1\dots\alpha_l\alpha_{l+1}\alpha_{l+2}\dots}\hat{u}^{\alpha_{l+1}}\hat{u}^{\alpha_{l+2}}\dots\,,
\end{equation}
\end{subequations}
where $\hat{u}^{\alpha}\equiv u^{\alpha}/\overline{u}$. As $d$ states the mass dimension of the field operator contracted with the controlling coefficient, the latter has mass dimension $4-d$. Therefore, $(\tilde{X}^{(d)})_{\alpha_1\dots\alpha_l}$ is a dimensionless quantity and also positively homogenous of degree 0 in the four-velocity. For consistency, the parameters $\xi^{(q)}_i$ of the $i$-th contribution at order $q$ (where $i\in \{1\dots q\}$) are dimensionless, as well, and depend on $d$ only. The results of these parameters for the spin-degenerate operators $\hat{a}_{\mu}$, $\hat{c}_{\mu}$, $\hat{e}$, and $\hat{m}$ are listed in \tabref{tab:results-parameters-lagrangians-a-e}.

The Lagrangian for $\hat{f}$ has a form different from all the other Lagrangians, as it does not involve odd powers in Lorentz violation:
\begin{subequations}
\label{eq:lagrangian-f-expansion-generic}
\begin{align}
L^{(d)}_{3,f}&=L_0^{(d)}\left\{1+(\tilde{f}^{(d)})^2\left[\xi_{1;f}^{(2)}+\xi^{(4)}_{1;f}(\tilde{f}^{(d)})^2+\xi^{(4)}_{2;f}(\tilde{f}^{(d)})_{\alpha}(\tilde{f}^{(d)})^{\alpha}\right.\right. \notag \\
&\phantom{{}={}L_0^{(d)}\Big\{}\left.\left.+\,\xi_{1;f}^{(6)}(\tilde{f}^{(d)})^4+\xi_{2;f}^{(6)}(\tilde{f}^{(d)})^2(\tilde{f}^{(d)})_{\alpha}(\tilde{f}^{(d)})^{\alpha}\right.\right. \notag \\
&\phantom{{}={}L_0^{(d)}\Big\{}\left.\left.+\,\xi_{3;f}^{(6)}\left((\tilde{f}^{(d)})_{\alpha}(\tilde{f}^{(d)})^{\alpha}\right)^2\right.\right. \notag \\
&\phantom{{}={}L_0^{(d)}\Big\{}\left.\left.+\,\xi_{4;f}^{(6)}\tilde{f}^{(d)}(\tilde{f}^{(d)})_{\alpha}(\tilde{f}^{(d)})^{\alpha\beta}(\tilde{f}^{(d)})_{\beta}\right]\right\}\,,
\end{align}
with
\begin{equation}
(\tilde{f}^{(d)})_{\alpha_1\dots\alpha_l}=m_{\psi}^{d-4}(f^{(d)})_{\alpha_1\dots\alpha_l\alpha_{l+1}\alpha_{l+2}\dots}\hat{u}^{\alpha_{l+1}}\hat{u}^{\alpha_{l+2}}\dots\,.
\end{equation}
\end{subequations}
\begin{table*}[t]
\centering
\setlength\extrarowheight{5pt}
\begin{tabular}{cc}
\toprule
Parameters & $\hat{f}^{(d)}$ \\
\midrule
$\xi^{(2)}_{1;f}$ & $\frac{1}{2}$ \\
$\xi^{(4)}_{1;f}$ & $-\frac{1}{8}(2d-7)^2$ \\
$\xi^{(4)}_{2;f}$ & $\frac{1}{2}(d-3)^2$ \\
$\xi^{(6)}_{1;f}$ & $\frac{1}{16}(2d-7)^4$ \\
$\xi^{(6)}_{2;f}$ & $-\frac{1}{4}(d-3)^2(2d-7)(4d-15)$ \\
$\xi^{(6)}_{3;f}$ & $\frac{1}{2}(d-3)^4$ \\
$\xi^{(6)}_{4;f}$ & $\frac{1}{2}(d-3)^3(d-4)$\\
\bottomrule
\end{tabular}
\caption{Parameters of the Lagrangian of \eqref{eq:lagrangian-f-expansion-generic} obtained for the spin-degenerate operator $\hat{f}$.}
\label{tab:results-parameters-lagrangians-f}
\end{table*}
The appropriate parameters can be taken from \tabref{tab:results-parameters-lagrangians-f}. Several remarks are in order. First, the parameters of the operators $\hat{a}^{(d)}_{\mu}$, $\hat{e}^{(d)}$ correspond to the negative ones of those for $\hat{c}^{(d)}_{\mu}$, $\hat{m}^{(d)}$ for odd powers of the controlling coefficients. The reason is that these operators jointly contribute to the dispersion equation in terms of a vector operator $\hat{\mathcal{V}}^{\mu}$ and a scalar operator $\hat{\mathcal{S}}$~\cite{Kostelecky:2013rta}. Second, the first-order contributions obtained here are in accordance with the results of \cite{Reis:2017ayl}. As the controlling coefficients and the four-velocities are the only quantities that can form objects without free Lorentz indices, antisymmetric coefficient choices cannot contribute at linear order, as mentioned at the beginning. Third, the parameters for $\hat{a}^{(d)}_{\mu}$ at quadratic and higher orders are equal to zero for $d=3$. This is expected, as the Lagrangian for the minimal $a$ coefficients is linear in $a^{(3)}_{\mu}$ (pseudo-Randers structure)~\cite{Kostelecky:2010hs}. Fourth, the parameters at leading order do not depend on the mass dimension. Furthermore, the parameters at second order are quadratic polynomials in $d$, whereas those at third order are even quartic polynomials in $d$. These polynomials do not factorize completely in rational numbers in contrast to the scalar field theory considered in~\cite{Edwards:2018lsn}. Fifth, the generalized correspondence between the $c$ and $f$ coefficients of \eqref{eq:correspondence-c-f} can be checked to map the Lagrangian of $\hat{c}_{\mu}$ to that of $\hat{f}$ by considering order by order. Doing this by hand can be tedious; an explicit demonstration to fourth order in $\hat{f}$ is shown in \ref{sec:map-c-f-coefficients}. Sixth, it is expected that there is a correspondence between the scalar $a$- and $c$-type coefficients and the $a$ and $c$ coefficients of the spin-degenerate fermion sector. In \cite{Edwards:2018lsn} this correspondence was found at leading order. At next-to-leading order we obtain (cf.~\ref{sec:connection-scalar-fermion}):
\begin{subequations}
\begin{align}
\label{eq:transformation-scalar-fermion}
(\tilde{a}^{(d)})\leftrightarrow \frac{1}{2}(\tilde{k}_a^{(d)})+\frac{1}{8}(\tilde{k}_a^{(d)})_{\alpha}(k_a^{(d)})^{\alpha}+\dots\,, \\[1ex]
\label{eq:transformation-scalar-fermion-c}
(\tilde{c}^{(d)})\leftrightarrow \frac{1}{2}(\tilde{k}_c^{(d)})-\frac{1}{8}(\tilde{k}_c^{(d)})_{\alpha}(k_c^{(d)})^{\alpha}+\dots\,,
\end{align}
\end{subequations}
with the quantities $\tilde{k}_{a,c}^{(d)}$ introduced in \cite{Edwards:2018lsn}. It is interesting that the transformation at second order in Lorentz violation is still independent of the mass dimension. Such a dependence is expected at third and higher orders.

Apart from these findings, at orders higher than linear it is not possible to express the Lagrangians in terms of the effective coefficients introduced in \cite{Kostelecky:2013rta}. The reason is that effective coefficients are linear-order concepts that do not apply to nonlinear contributions. We can also compare the result for the fermion $\hat{c}_{\mu}$ operator to Eq.~(12) of~\cite{Edwards:2018lsn} stating the third-order Lagrangian for the scalar field theory including $c$-type coefficients. As we work in four spacetime dimensions, $n=4$ has to be inserted into Eq.~(12). By doing so, we observe two differences. First, comparing $\mathcal{D}(p)$ of the modified fermion dispersion equation to the left-hand side of their Eq.~(3) reveals the correspondence
$\hat{c}_{\mu\nu}\leftrightarrow (\hat{k}_c)_{\mu\nu}/2$ (cf.~also the first term of \eqref{eq:transformation-scalar-fermion-c}). Hence, all parameters at order $q$ in Eq.~(12) of \cite{Edwards:2018lsn} involve additional factors of $2^{-q}$ in comparison to the parameters obtained for fermions. This difference is relatively easy to understand, as its connection to spin-1/2 is evident. The second one is quite intransparent, though. We observe that the parameters $\xi^{(q)}_{\geq 2}$ extensively differ from their counterparts in Eq.~(12) of \cite{Edwards:2018lsn}. These discrepancies are spin-induced, as well, but much more involved than the additional factors of 1/2 mentioned before. The latter spin effects appear for a subset of parameters at quadratic and higher orders.

\section{Conclusions}
\label{sec:conclusions}

This article was dedicated to deriving classical Lagrangians associated with the nonminimal, spin-degenerate fermion operators of the SME at third order in Lorentz violation. These results were obtained for operators of arbitrary mass dimension. The base was a perturbative, iterative technique developed in the recent work \cite{Edwards:2018lsn}. The current paper demonstrated how the latter method can be applied to modified Dirac fermions --- at least those described by degenerate dispersion laws. The classical Lagrangians obtained share some of the properties of those derived for a scalar field theory in \cite{Edwards:2018lsn}. However, there are also essential differences caused by spin effects.

A reasonable next step would be to adopt the method to treat the spin-nondegenerate fermion operators $\hat{b}_{\mu}$, $\hat{d}_{\mu}$, $\hat{H}_{\mu\nu}$, and $\hat{g}_{\mu\nu}$. This task is expected to require certain adaptations of the method, as the dispersion equation for these operators is not simply the square of a quadratic equation. Another direction to be pursued in the future is to promote the resulting classical Lagrangians to Finsler structures and to investigate their mathematical properties, as carried out in \cite{Edwards:2018lsn}. Since the generic form of the Lagrangians for the spin-degenerate operators found here is similar to that of the Lagrangians in \cite{Edwards:2018lsn}, the Finsler structures associated are also expected to have similar characteristics.

\section*{Acknowledgments}

The author greatly acknowledges the hospitality and financial support of the Indiana University Center for Spacetime Symmetries (IUCSS) where this research was carried out. The author is also indebted to V.A~Kosteleck\'{y} for encouraging this line of research and thanks V.A.~Kosteleck\'{y} as well as B.~Edwards for discussions on the subject. Furthermore, the scientific mission at the IUCSS was financially supported by the Brazilian agencies CNPq and FAPEMA with grant no. 421566/2016-7 and 01149/17, respectively.

\appendix
\section{Computation of four-velocities}
\label{sec:computation-four-velocities}

In the current section we demonstrate briefly how to obtain the four-velocities of Eqs.~(\ref{eq:four-velocities-cases}) for certain characteristic cases. The general dispersion equation for the nonminimal spin-degenerate operators can be written in the form of \eqref{eq:dispersion-equation-generic-minimal-spin-degenerate}. We consider the operators $\hat{a}_{\mu}$, $\hat{c}_{\mu}$, and $\hat{e}$. Each operator covers one of the three particular cases $\kappa(p)\neq 0$, $\Omega^{\mu\nu}(p)\neq \eta^{\mu\nu}$, and $\mu(p)\neq m_{\psi}$. The calculation for $\hat{m}$ and $\hat{f}$ is analog to that for $\hat{e}$ and $\hat{c}_{\mu}$, respectively.

\subsection{Operator $\hat{a}_{\mu}$}

The dispersion equation for the operator $\hat{a}_{\mu}$ has the form
\begin{align}
\label{eq:dispersion-equation-a}
m_{\psi}^2&=\left(p_0-(a^{(d)})_0^{\phantom{0}\diamond}\right)\Omega^{00}\left(p_0-(a^{(d)})_0^{\phantom{0}\diamond}\right) \notag \\
&\phantom{{}={}}+2\left(p_i-(a^{(d)})_i^{\phantom{i}\diamond}\right)\Omega^{i0}\left(p_0-(a^{(d)})_0^{\phantom{0}\diamond}\right) \notag \\
&\phantom{{}={}}+\left(p_k-(a^{(d)})_k^{\phantom{k}\diamond}\right)\Omega^{kl}\left(p_l-(a^{(d)})_l^{\phantom{l}\diamond}\right)\,,
\end{align}
with $\Omega^{\mu\nu}=\eta^{\mu\nu}$. Differentiation with respect to $p_j$ gives
\begin{align}
0&=\left(\frac{\partial p_0}{\partial p_j}-\frac{\partial (a^{(d)})_0^{\phantom{0}\diamond}}{\partial p_j}\right)\Omega^{00}\left(p_0-(a^{(d)})_0^{\phantom{0}\diamond}\right) \notag \\
&\phantom{{}={}}+\left(\delta_{ij}-\frac{\partial(a^{(d)})_i^{\phantom{i}\diamond}}{\partial p_j}\right)\Omega^{i0}\left(p_0-(a^{(d)})_0^{\phantom{0}\diamond}\right) \notag \displaybreak[0]\\
&\phantom{{}={}}+\left(p_i-(a^{(d)})_i^{\phantom{i}\diamond}\right)\Omega^{i0}\left(\frac{\partial p_0}{\partial p_j}-\frac{\partial (a^{(d)})_0^{\phantom{0}\diamond}}{\partial p_j}\right) \notag \\
&\phantom{{}={}}+\left(\delta_{kj}-\frac{\partial(a^{(d)})_k^{\phantom{k}\diamond}}{\partial p_j}\right)\Omega^{kl}\left(p_l-(a^{(d)})_l^{\phantom{l}\diamond}\right) \notag \displaybreak[0]\\
&=2\left(\frac{\partial p_0}{\partial p_j}-\frac{\partial (a^{(d)})_0^{\phantom{0}\diamond}}{\partial p_j}\right)\Omega^{0\nu}\left(p_{\nu}-(a^{(d)})_{\nu}^{\phantom{\nu}\diamond}\right) \notag \\
&\phantom{{}={}}+2\left(\delta_{ij}-\frac{\partial (a^{(d)})_i^{\phantom{i}\diamond}}{\partial p_j}\right)\Omega^{i\nu}\left(p_{\nu}-(a^{(d)})_{\nu}^{\phantom{\nu}\diamond}\right)\,.
\end{align}
As $(a^{(d)})^{\mu}$ is positively homogeneous of degree $d-2$ in the four-momentum, we obtain
\begin{equation}
\label{eq:homogeneity-equation-a}
p_j\frac{\partial (a^{(d)})^{\mu\diamond}}{\partial p_j}=(d-3)\left[(a^{(d)})^{\mu\diamond}-\frac{p\cdot u}{u^0}(a^{(d)})^{\mu0\diamond}\right]\,,
\end{equation}
where Eqs.~(\ref{eq:mapping-equation-2-3-4}) were also employed. Using the latter and Eqs.~(\ref{eq:mapping-equation-2-3-4}) in \eqref{eq:dispersion-equation-a} provides
\begin{align}
0&=\left(-p_j\frac{u^j}{u^0}\right)\Omega^{0\nu}p_{\nu}+p_j\Omega^{j\nu}p_{\nu}-p_j\frac{\partial(a^{(d)})_{\mu}^{\phantom{\mu}\diamond}}{\partial p_j}\Omega^{\mu\nu}p_{\nu} \notag \\
&\phantom{{}={}}-\left(-p_j\frac{u^j}{u^0}\right)\Omega^{0\nu}(a^{(d)})_{\nu}^{\phantom{\nu}\diamond}-p_j\Omega^{j\nu}(a^{(d)})_{\nu}^{\phantom{\nu}\diamond} \notag \\
&\phantom{{}={}}+p_j\frac{\partial (a^{(d)})_{\mu}^{\phantom{\mu}\diamond}}{\partial p_j}\Omega^{\mu\nu}(a^{(d)})_{\nu}^{\phantom{\nu}\diamond}\,.
\end{align}
Setting $\Omega^{\mu\nu}=\eta^{\mu\nu}$ and inserting \eqref{eq:homogeneity-equation-a} as well as the original dispersion equation again results in
\begin{subequations}
\begin{align}
u^0&\left\{m_{\psi}^2+p_0(a^{(d)})^{0\diamond}+p_{\mu}(a^{(d)})^{\mu\diamond}-(a^{(d)})_{\mu}^{\phantom{\mu}\diamond}(a^{(d)})^{\mu\diamond}\right. \notag \\
&\phantom{{}={}}\left.-(d-3)\left[(a^{(d)})_{\mu}^{\phantom{\mu}\diamond}-\frac{p\cdot u}{u^0}(a^{(d)})_{0\mu}^{\phantom{0\mu}\diamond}\right]\left(p^{\mu}-(a^{(d)})^{\mu\diamond}\right)\right\} \notag \\
&=-Lp^0-p_ju^j(a^{(d)})^{0\diamond}\,.
\end{align}
Further simplification gives
\begin{align}
u^0&\left[m_{\psi}^2+p_{\mu}(a^{(d)})^{\mu\diamond}-(a^{(d)})_{\mu}^{\phantom{\mu}\diamond}(a^{(d)})^{\mu\diamond}\right. \notag \\
&\phantom{{}={}}\left.-(d-3)(a^{(d)})^{\mu\diamond}\left(p_{\mu}-(a^{(d)})_{\mu}^{\phantom{\mu}\diamond}\right)\right] \notag \\
&=-L\left(p^0-(a^{(d)})^{0\diamond}\right) \notag \\
&\phantom{{}={}}-(d-3)p\cdot u (a^{(d)})^{0\mu\diamond}\left(p_{\mu}-(a^{(d)})_{\mu}^{\phantom{\mu}\diamond}\right)\,.
\end{align}
\end{subequations}
Finally, after applying \eqref{eq:homogeneity-equation-a} once more, the resulting equation can be readily solved for the zeroth component of the four-velocity:
\begin{equation}
u^0=-L\frac{p^0-(a^{(d)})^{0\diamond}-(d-3)(a^{(d)})^{0\mu\diamond}\left(p_{\mu}-(a^{(d)})_{\mu}^{\phantom{\mu}\diamond}\right)}{m_{\psi}^2-(d-4)(a^{(d)})^{\mu\diamond}\left(p_{\mu}-(a^{(d)})_{\mu}^{\phantom{\mu}\diamond}\right)}\,.
\end{equation}
Moreover, we obtain
\begin{align}
u^jp_j&=-u^0p_0-L \notag \\
&=L\left[p^0p_0-(a^{(d)})^{0\diamond}p_0-(a^{(d)})^{\mu\diamond}\left(p_{\mu}-(a^{(d)})_{\mu}^{\phantom{\mu}\diamond}\right)\right. \notag \\
&\phantom{{}={}L\Big\{}\left.-\,m_{\psi}^2+(d-3)(a^{(d)})^{j\mu\diamond}p_j\left(p_{\mu}-(a^{(d)})_{\mu}^{\phantom{\mu}\diamond}\right)\right] \notag \\
&\phantom{{}={}}\times\left\{m_{\psi}^2-(d-4)(a^{(d)})^{\mu\diamond}\left(p_{\mu}-(a^{(d)})_{\mu}^{\phantom{\mu}\diamond}\right)\right\}^{-1} \notag \\
&=-L\frac{p^jp_j-(a^{(d)})^{j\diamond}p_j-(d-3)(a^{(d)})^{j\mu\diamond}p_j\left(p_{\mu}-(a^{(d)})_{\mu}^{\phantom{\mu}\diamond}\right)}{m_{\psi}^2-(d-4)(a^{(d)})^{\mu\diamond}\left(p_{\mu}-(a^{(d)})_{\mu}^{\phantom{\mu}\diamond}\right)}\,,
\end{align}
where we used
\begin{equation}
p^{\mu}\left(p_{\mu}-(a^{(d)})_{\mu}^{\phantom{\mu}\diamond}\right)=m_{\psi}^2+(a^{(d)})^{\mu\diamond}\left(p_{\mu}-(a^{(d)})_{\mu}^{\phantom{\mu}\diamond}\right)\,,
\end{equation}
originating from \eqref{eq:dispersion-equation-a}. Hence, we can read off the spatial components of the four-velocity:
\begin{equation}
u^j=-L\frac{p^j-(a^{(d)})^{j\diamond}-(d-3)(a^{(d)})^{j\mu\diamond}\left(p_{\mu}-(a^{(d)})_{\mu}^{\phantom{\mu}\diamond}\right)}{m_{\psi}^2-(d-4)(a^{(d)})^{\mu\diamond}\left(p_{\mu}-(a^{(d)})_{\mu}^{\phantom{\mu}\diamond}\right)}\,.
\end{equation}
These results confirm \eqref{eq:four-velocity-a}.

\subsection{Operador $\hat{c}_{\mu}$}

The dispersion equation for the nonminimal operator $\hat{c}_{\mu}$ is cast into the form
\begin{subequations}
\begin{align}
m_{\psi}^2&=p_{\mu}\Omega^{\mu\nu}p_{\nu} \notag \\
&=p_0\Omega^{00}p_0+p_i\Omega^{i0}p_0+p_0\Omega^{0i}p_i+p_k\Omega^{kl}p_l\,,
\end{align}
where
\begin{equation}
\Omega^{\mu\nu}=\Omega^{\mu\nu}(p)=\eta^{\mu\nu}+2(c^{(d)})^{\mu\nu\diamond}+(c^{(d)})^{\mu\kappa\diamond}(c^{(d)})_{\kappa}^{\phantom{\kappa}\nu\diamond}\,.
\end{equation}
\end{subequations}
Differentiation with respect to $p_j$ gives
\begin{subequations}
\begin{align}
0&=2\frac{\partial p_0}{\partial p_j}\Omega^{00}p_0+2\Omega^{j0}p_0+2p_i\Omega^{i0}\frac{\partial p_0}{\partial p_j}+2\Omega^{jl}p_l \notag \\
&\phantom{{}={}}+p_{\mu}\frac{\partial\Omega^{\mu\nu}}{\partial p_j}p_{\nu}\,,
\end{align}
where we use Eqs.~(\ref{eq:mapping-equation-2-3-4}) to obtain
\begin{equation}
0=u^0\Omega^{j\nu}p_{\nu}-u^j\Omega^{0\nu}p_{\nu}+\frac{u^0}{2}p_{\mu}\frac{\partial\Omega^{\mu\nu}}{\partial p_j}p_{\nu}\,.
\end{equation}
\end{subequations}
Contracting the latter with $p_j$ leads to
\begin{equation}
\label{eq:contracted-derivative-disp-eq-c}
0=u^0p_j\Omega^{j\nu}p_{\nu}-p_ju^j\Omega^{0\nu}p_{\nu}+\frac{u^0}{2}p_{\mu}p_j\frac{\partial\Omega^{\mu\nu}}{\partial p_j}p_{\nu}\,.
\end{equation}
As $c^{(d)\diamond}$ is positively homogeneous of degree $d-3$ in the momentum, its derivative for $p_j$ contracted with $p_j$ is
\begin{equation}
p_j\frac{\partial(c^{(d)})^{\diamond}}{\partial p_j}=(d-2)\left[(c^{(d)})^{\diamond}-\frac{p\cdot u}{u^0}(c^{(d)})^{0\diamond}\right]\,.
\end{equation}
We use this result to compute the final term in \eqref{eq:contracted-derivative-disp-eq-c}:
\begin{align}
p_j\frac{\partial\Omega^{\mu\nu}}{\partial p_j}&=2(d-4)\left[(c^{(d)})^{\mu\nu\diamond}-\frac{p\cdot u}{u^0}(c^{(d)})^{\mu\nu0\diamond}\right] \notag \\
&\phantom{{}={}}+(d-4)\left\{\left[(c^{(d)})^{\mu\kappa\diamond}-\frac{p\cdot u}{u^0}(c^{(d)})^{\mu\kappa0\diamond}\right](c^{(d)})_{\kappa}^{\phantom{\kappa}\nu\diamond}\right. \notag \\
&\phantom{{}={}(d)}\left.+\,(c^{(d)})^{\mu\kappa\diamond}\left[(c^{(d)})_{\kappa}^{\phantom{\kappa}\nu\diamond}-\frac{p\cdot u}{u^0}(c^{(d)})_{\kappa}^{\phantom{\kappa}\nu0\diamond}\right]\right\}\,.
\end{align}
Employing the latter finding as well as \eqref{eq:mapping-equation-1} again leads to
\begin{align}
0&=u^0(m_{\psi}^2-p_0\Omega^{0\nu}p_{\nu})-p_ju^j\Omega^{0\nu}p_{\nu} \notag \\
&\phantom{{}={}}+(d-4)\left[u^0(c^{(d)})^{\diamond}+L(c^{(d)})^{0\diamond}\right] \notag \\
&\phantom{{}={}}+\frac{d-4}{2}\left\{\left[u^0(c^{(d)})^{\kappa\diamond}+L(c^{(d)})^{\kappa0\diamond}\right](c^{(d)})_{\kappa}^{\phantom{\kappa}\diamond}\right. \notag \\
&\phantom{{}={}}\left.\hspace{1.5cm}+\,(c^{(d)})^{\kappa\diamond}\left[u^0(c^{(d)})_{\kappa}^{\phantom{\kappa}\diamond}+L(c^{(d)})_{\kappa}^{\phantom{\kappa}0\diamond}\right]\right\}\,.
\end{align}
In addition, inserting
\begin{equation}
p_{\mu}\Omega^{\mu\nu}p_{\nu}=p^2+2(c^{(d)})^{\diamond}+(c^{(d)})^{\kappa\diamond}(c^{(d)})_{\kappa}^{\phantom{\kappa}\diamond}=m_{\psi}^2\,,
\end{equation}
provides
\begin{align}
0&=u^0\left\{m_{\psi}^2+(d-4)\left[(c^{(d)})^{\diamond}+(c^{(d)})^{\kappa\diamond}(c^{(d)})_{\kappa}^{\phantom{\kappa}\diamond}\right]\right\} \notag \\
&\phantom{{}={}}+L\left\{\eta^{0\nu}p_{\nu}+(d-2)(c^{(d)})^{0\diamond}\right. \notag \\
&\phantom{{}={}+L\{}\left.+\,(d-3)(c^{(d)})^{0\kappa\diamond}(c^{(d)})_{\kappa}^{\phantom{\kappa}\diamond}\right\}\,.
\end{align}
Finally, this relation is solved for $u^0$:
\begin{equation}
u^0=-L\frac{p^0+(d-2)(c^{(d)})^{0\diamond}+(d-3)(c^{(d)})^{0\kappa\diamond}(c^{(d)})_{\kappa}^{\phantom{\kappa}\diamond}}{m_{\psi}^2+(d-4)\left[(c^{(d)})^{\diamond}+(c^{(d)})^{\kappa\diamond}(c^{(d)})_{\kappa}^{\phantom{\kappa}\diamond}\right]}\,.
\end{equation}
Based on \eqref{eq:mapping-equation-5}, we calculate the spatial components of the four-velocity explicitly:
\begin{align}
u^jp_j&=-u^0p_0-L \notag \\
&=L\left\{p^0p_0+(d-2)(c^{(d)})^{0\diamond}p_0+(d-3)(c^{(d)})^{0\kappa\diamond}(c^{(d)})_{\kappa}^{\phantom{\kappa}\diamond}p_0\right. \notag \\
&\phantom{{}={}L\{}\left.-m_{\psi}^2-(d-4)\left[(c^{(d)})^{\diamond}+(c^{(d)})^{\kappa\diamond}(c^{(d)})_{\kappa}^{\phantom{\kappa}\diamond}\right]\right\} \notag \\
&\phantom{{}={}}\times \left\{m_{\psi}^2+(d-4)\left[(c^{(d)})^{\diamond}+(c^{(d)})^{\kappa\diamond}(c^{(d)})_{\kappa}^{\phantom{\kappa}\diamond}\right]\right\}^{-1} \notag \\
&=L\left\{-p^jp_j+(d-2)(c^{(d)})^{0\diamond}p_0+(d-3)(c^{(d)})^{0\kappa\diamond}(c^{(d)})_{\kappa}^{\phantom{\kappa}\diamond}p_0\right. \notag \\
&\phantom{{}={}L\}}\left.-(d-2)(c^{(d)})^{\diamond}-(d-3)(c^{(d)})^{\kappa\diamond}(c^{(d)})_{\kappa}^{\phantom{\kappa}\diamond}\right\} \notag \\
&\phantom{{}={}}\times\left\{m_{\psi}^2+(d-4)\left[(c^{(d)})^{\diamond}+(c^{(d)})^{\kappa\diamond}(c^{(d)})_{\kappa}^{\phantom{\kappa}\diamond}\right]\right\}^{-1} \notag \\
&=-L\frac{p^jp_j+(d-2)(c^{(d)})^{j\diamond}p_j+(d-3)(c^{(d)})^{j\kappa\diamond}(c^{(d)})_{\kappa}^{\phantom{\kappa}\diamond}p_j}{m_{\psi}^2+(d-4)\left[(c^{(d)})^{\diamond}+(c^{(d)})^{\kappa\diamond}(c^{(d)})_{\kappa}^{\phantom{\kappa}\diamond}\right]}\,.
\end{align}
Comparing both sides of this equation, we can immediately identify the spatial components as follows:
\begin{equation}
u^j=-L\frac{p^j+(d-2)(c^{(d)})^{j\diamond}+(d-3)(c^{(d)})^{j\kappa\diamond}(c^{(d)})_{\kappa}^{\phantom{\kappa}\diamond}}{m_{\psi}^2+(d-4)\left[(c^{(d)})^{\diamond}+(c^{(d)})^{\kappa\diamond}(c^{(d)})_{\kappa}^{\phantom{\kappa}\diamond}\right]}\,.
\end{equation}
Both $u^0$ and $u^j$ are in accordance with \eqref{eq:four-velocity-c}.

\subsection{Operator $\hat{e}$}

The dispersion equation for the operator $\hat{e}$ has the form
\begin{subequations}
\label{eq:dispersion-equation-e}
\begin{align}
\mu^2&=p_{\mu}\Omega^{\mu\nu}p_{\nu} \notag \\
&=p_0\Omega^{00}p_0+p_i\Omega^{i0}p_0+p_0\Omega^{0i}p_i+p_k\Omega^{kl}p_l\,,
\end{align}
with
\begin{equation}
\mu=\mu(p)=m_{\psi}-(e^{(d)})^{\diamond}\,,\quad\Omega^{\mu\nu}=\eta^{\mu\nu}\,.
\end{equation}
\end{subequations}
Differentiation with respect to $p_j$ provides
\begin{subequations}
\begin{equation}
\mu\frac{\partial \mu}{\partial p_j}=\frac{\partial p_0}{\partial p_j}\Omega^{00}p_0+\Omega^{j0}p_0+p_i\Omega^{i0}\frac{\partial p_0}{\partial p_j}+\Omega^{jl}p_l\,,
\end{equation}
and
\begin{equation}
\label{eq:contracted-derivative-disp-eq-e}
u^0\mu\frac{\partial\mu}{\partial p_j}=u^0\Omega^{j\nu}p_{\nu}-u^j\Omega^{0\nu}p_{\nu}\,.
\end{equation}
\end{subequations}
As $(e^{(d)})^{\diamond}$ is positively homogeneous of degree $d-3$ in the four-momentum, we obtain
\begin{equation}
p_j\frac{\partial(e^{(d)})^{\diamond}}{\partial p_j}=(d-3)\left[(e^{(d)})^{\diamond}-\frac{p\cdot u}{u^0}(e^{(d)})^{0\diamond}\right]\,.
\end{equation}
Inserting the latter result into \eqref{eq:contracted-derivative-disp-eq-e} leads to
\begin{align}
-u^0\mu&(d-3)\left[(e^{(d)})^{\diamond}-\frac{p\cdot u}{u^0}(e^{(d)})^{0\diamond}\right] \notag \\
&=u^0[\mu^2-p_0\Omega^{0\nu}p_{\nu}]-p_ju^j\Omega^{0\nu}p_{\nu}\,.
\end{align}
Employing $\Omega^{\mu\nu}=\eta^{\mu\nu}$ and \eqref{eq:mapping-equation-5} results in
\begin{align}
u^0\mu&\left[(d-4)(e^{(d)})^{\diamond}+m_{\psi}\right] \notag \\
&=-L\left[p^0+(d-3)\mu(e^{(d)})^{0\diamond}\right]\,.
\end{align}
Finally, solving this relation for $u^0$ gives
\begin{equation}
u^0=-L\frac{p^0+(d-3)\left(m_{\psi}-(e^{(d)})^{\diamond}\right)(e^{(d)})^{0\diamond}}{\left[m_{\psi}+(d-4)(e^{(d)})^{\diamond}\right]\left(m_{\psi}-(e^{(d)})^{\diamond}\right)}\,.
\end{equation}
Moreover,
\begin{align}
u^jp_j&=-u^0p_0-L \notag \\
&=L\left[p^0p_0-\left(m_{\psi}-(e^{(d)})^{\diamond}\right)^2\right. \notag \\
&\phantom{{}={}L\Big[}\left.-\,(d-3)\left(m_{\psi}-(e^{(d)})^{\diamond}\right)(e^{(d)})^{j\diamond}p_j\right] \notag \\
&\phantom{{}={}}\times\left\{\left[m_{\psi}+(d-4)(e^{(d)})^{\diamond}\right]\left(m_{\psi}-(e^{(d)})^{\diamond}\right)\right\}^{-1} \notag \\
&=-L\frac{p^jp_j+(d-3)\left(m_{\psi}-(e^{(d)})^{\diamond}\right)(e^{(d)})^{j\diamond}p_j}{\left[m_{\psi}+(d-4)(e^{(d)})^{\diamond}\right]\left(m_{\psi}-(e^{(d)})^{\diamond}\right)}\,,
\end{align}
where \eqref{eq:dispersion-equation-e} was used again. Hence,
\begin{equation}
u^j=-L\frac{p^j+(d-3)\left(m_{\psi}-(e^{(d)})^{\diamond}\right)(e^{(d)})^{j\diamond}}{\left[m_{\psi}+(d-4)(e^{(d)})^{\diamond}\right]\left(m_{\psi}-(e^{(d)})^{\diamond}\right)}\,,
\end{equation}
confirming \eqref{eq:four-velocity-e}.

\section{Map between $c$ and $f$ coefficients}
\label{sec:map-c-f-coefficients}

The correspondence (\ref{eq:correspondence-c-f}) between the operators $\hat{c}_{\mu}$ and $\hat{f}$ valid at the level of field theory can be checked as follows. The corresponding dispersion equations are obtained from Eq.~(39) of~\cite{Kostelecky:2013rta}:
\begin{subequations}
\begin{align}
\label{eq:dispersion-equation-c}
0&=p^2-m_{\psi}^2+2(c^{(d)})^{\diamond}+(c^{(d)})_{\alpha}^{\phantom{\alpha}\diamond}(c^{(d)})^{\alpha\diamond}\,, \displaybreak[0]\\[1ex]
\label{eq:dispersion-equation-f}
0&=p^2-m_{\psi}^2-[(f^{(\overline{d})})^{\diamond}]^2\,.
\end{align}
\end{subequations}
We compute the Lorentz-violating contributions of \eqref{eq:dispersion-equation-c} in replacing the $c$ coefficients by the $f$ coefficients via \eqref{eq:correspondence-c-f}:
\begin{subequations}
\begin{align}
(c^{(d)})^{\diamond}&=\frac{[(f^{(\overline{d})})^{\diamond}]^2}{\Theta}\left[\sqrt{1-\Theta}-1\right]\,, \displaybreak[0]\\[1ex]
(c^{(d)})_{\alpha}^{\phantom{\alpha}\diamond}(c^{(d)})^{\alpha\diamond}&=\frac{\Theta[(f^{(\overline{d})})^{\diamond}]^2}{\Theta^2}\left[\sqrt{1-\Theta}-1\right]^2 \notag \\
&=\frac{[(f^{(\overline{d})})^{\diamond}]^2}{\Theta}\left[2-\Theta-2\sqrt{1-\Theta}\right]\,.
\end{align}
\end{subequations}
Adding the appropriate linear combination of both expressions leads to
\begin{align}
2(c^{(d)})^{\diamond}&+(c^{(d)})_{\alpha}^{\phantom{\alpha}\diamond}(c^{(d)})^{\alpha\diamond} \notag \\
&=\frac{[(f^{(\overline{d})})^{\diamond}]^2}{\Theta}\left[2\sqrt{1-\Theta}-2+2-\Theta-2\sqrt{1-\Theta}\right] \notag \\
&=-\frac{[(f^{(\overline{d})})^{\diamond}]^2}{\Theta}\Theta=-[(f^{(\overline{d})})^{\diamond}]^2\,.
\end{align}
Hence, the latter result corresponds to the Lorentz-violating piece in \eqref{eq:dispersion-equation-f} demonstrating the validity of the map.
As the $c$ and $f$ coefficients do not mix with each other, the total dispersion equation can also be expressed as
\begin{subequations}
\begin{align}
0&=p_{\mu}(\eta^{\mu\nu}+\Theta^{\mu\nu})p_{\nu}-m_{\psi}^2\,, \\[2ex]
\Theta^{\mu\nu}=\Theta^{\mu\nu}(p)&=2(c^{(d)})^{\mu\nu\diamond}+(c^{(d)})^{\mu\phantom{\alpha}\diamond}_{\phantom{\mu}\alpha}(c^{(d)})^{\alpha\nu\diamond} \notag \\
&\phantom{{}={}}-(f^{(\overline{d})})^{\mu\diamond}(f^{(\overline{d})})^{\nu\diamond}\,.
\end{align}
\end{subequations}
Thus, with nonzero $c$ coefficients present, new coefficients $c'$ can be defined by a field redefinition transforming all $f$ coefficients into the $c$ sector:
\begin{equation}
\label{eq:redefinition-c-coefficients}
(c'^{(d)})^{\mu\phantom{\nu}\diamond}_{\phantom{\mu}\nu}=\sqrt{\delta^{\mu}_{\phantom{\mu}\nu}+\Theta^{\mu}_{\phantom{\mu}\nu}}-\delta^{\mu}_{\phantom{\mu}\nu}\,.
\end{equation}
The latter must be treated as an infinite matrix series. It is the generalization of Eq.~(11) in \cite{Kostelecky:2010hs} for the nonminimal SME. Introducing the trace $\Theta\equiv\Theta_{\alpha}^{\phantom{\alpha}\alpha}$ and setting all $c$ coefficients on the right-hand side of \eqref{eq:redefinition-c-coefficients} equal to zero, this relation can be also formulated as follows:
\begin{align}
(c^{(d)})^{\mu\phantom{\nu}\diamond}_{\phantom{\mu}\nu}&=\sqrt{\delta^{\mu}_{\phantom{\mu}\nu}-(f^{(\overline{d})})^{\mu\diamond}(f^{(\overline{d})})_{\nu}^{\phantom{\nu}\diamond}}-\delta^{\mu}_{\phantom{\mu}\nu} \notag \displaybreak[0]\\
&=-\frac{1}{2}(f^{(\overline{d})})^{\mu\diamond}(f^{(\overline{d})})_{\nu}^{\phantom{\nu}\diamond}-\frac{1}{8}(f^{(\overline{d})})^{\mu\diamond}\Theta(f^{(\overline{d})})_{\nu}^{\phantom{\nu}\diamond} \notag \displaybreak[0]\\
&\phantom{{}={}}-\frac{1}{16}(f^{(\overline{d})})^{\mu\diamond}\Theta^2(f^{(\overline{d})})_{\nu}^{\phantom{\nu}\diamond}-\dots \notag \displaybreak[0]\\
&=-(f^{(\overline{d})})^{\mu\diamond}(f^{(\overline{d})})_{\nu}^{\phantom{\nu}\diamond}\left[\frac{1}{2}+\frac{\Theta}{8}+\frac{\Theta^2}{16}+\dots\right] \notag \displaybreak[0]\\
&=\frac{(f^{(\overline{d})})^{\mu\diamond}(f^{(\overline{d})})_{\nu}^{\phantom{\nu}\diamond}}{\Theta}\left[\sqrt{1-\Theta}-1\right]\,,
\end{align}
which corresponds to \eqref{eq:correspondence-c-f}. This map is expected to have an equivalent in the classical regime. Hence, the parameters $\xi^{(q)}_{i;c}$ of \tabref{tab:results-parameters-lagrangians-a-e} and $\xi^{(2q)}_{i;f}$ of \tabref{tab:results-parameters-lagrangians-f} should be related to each other. This connection is relatively easy to see at linear order in Lorentz violation (cf.~\cite{Reis:2017ayl} for a discussion). However, at higher orders in Lorentz violation one must carefully keep track of all terms at the particular order considered. In what follows, the connection between the parameters $\xi^{(2)}_{i;c}$ and $\xi^{(4)}_{i;f}$ shall be demonstrated. The contribution at first order in the $c$ coefficients results in a term at second order and another one at quartic order in the $f$ coefficients:
\begin{equation}
-(\tilde{c}^{(d)})\mapsto \frac{1}{2}(\tilde{f}^{(\overline{d})})^2+\frac{1}{8}(\tilde{f}^{(\overline{d})})^2(\tilde{f}^{(\overline{d})})_{\alpha}(\tilde{f}^{(\overline{d})})^{\alpha}+\dots\,.
\end{equation}
The first term confirms $\xi_{1;f}^{(2)}$ in \tabref{tab:results-parameters-lagrangians-f}. The contributions at second order in $c$ provide terms at quartic order in $f$:
\begin{subequations}
\begin{equation}
-\frac{1}{2}(d-3)^2(\tilde{c}^{(d)})^2\mapsto -\frac{1}{8}(2\overline{d}-7)^2(\tilde{f}^{(\overline{d})})^4+\dots\,,
\end{equation}
and
\begin{align}
\frac{1}{2}&(d-1)(d-3)(\tilde{c}^{(d)})_{\alpha}(\tilde{c}^{(d)})^{\alpha} \notag \\
&\mapsto \frac{1}{8}(2\overline{d}-5)(2\overline{d}-7)(\tilde{f}^{(\overline{d})})^2(\tilde{f}^{(\overline{d})})_{\alpha}(\tilde{f}^{(\overline{d})})^{\alpha}+\dots\,.
\end{align}
\end{subequations}
The first quartic term is in accordance with $\xi_{1;f}^{(4)}$. Adding the remaining matching terms at fourth order leads to
\begin{align}
\frac{1}{8}&\left[1+(2\overline{d}-5)(2\overline{d}-7)\right](\tilde{f}^{(\overline{d})})^2(\tilde{f}^{(\overline{d})})_{\alpha}(\tilde{f}^{(\overline{d})})^{\alpha} \notag \\
&\mapsto\frac{1}{2}(\overline{d}-3)^2(\tilde{f}^{(\overline{d})})^2(\tilde{f}^{(\overline{d})})_{\alpha}(\tilde{f}^{(\overline{d})})^{\alpha}+\dots\,,
\end{align}
which is equal to $\xi_{2;f}^{(4)}$ in \tabref{tab:results-parameters-lagrangians-f}. It is possible to proceed in a similar manner for higher orders, although the computations become more tedious.

\section{Connection between scalar and fermion Lagrangians}
\label{sec:connection-scalar-fermion}

We consider the Lagrangian (\ref{eq:lagrangian-expansion-generic}) obtained for the operator $\hat{a}_{\mu}$ at second order in Lorentz violation and apply the transformation of \eqref{eq:transformation-scalar-fermion}:
\begin{align}
L_{2;a}^{(d)}&=L_0^{(d)}\left[1+\tilde{a}^{(d)}-\frac{1}{2}(d-3)^2(\tilde{a}^{(d)})^2\right. \notag \\
&\phantom{{}={}L_0^{(d)}\Big[}\left.+\,\frac{1}{2}(d-1)(d-3)(\tilde{a}^{(d)})_{\alpha}(\tilde{a}^{(d)})^{\alpha}+\dots\right] \notag \displaybreak[0]\\
&\mapsto L_0^{(d)}\left\{1+\frac{1}{2}(\tilde{k}_a^{(d)})-\frac{1}{8}(d-3)^2(\tilde{k}_a^{(d)})^2\right. \notag \\
&\phantom{{}={}L_0^{(d)}\Bigg\{}\left.+\,\frac{1}{8}\left[(d-1)(d-3)+1\right](\tilde{k}^{(d)}_a)_{\alpha}(\tilde{k}^{(d)}_a)^{\alpha}+\dots\right\} \notag \displaybreak[0]\\
&=L_0^{(d)}\left[1+\frac{1}{2}(\tilde{k}_a^{(d)})-\frac{1}{8}(d-3)^2(\tilde{k}_a^{(d)})^2\right. \notag \\
&\phantom{{}={}L_0^{(d)}\Big[}\left.+\,\frac{1}{8}(d-2)^2(\tilde{k}^{(d)}_a)_{\alpha}(\tilde{k}^{(d)}_a)^{\alpha}+\dots\right]\,,
\end{align}
which corresponds to Eq.~(12) of \cite{Edwards:2018lsn} for $n=4$ at second order. The demonstration works in the same manner for the operator~$\hat{c}_{\mu}$.

\end{document}